\documentclass[preprint,aps]{revtex4}
\usepackage{epsf}
\usepackage{graphicx} 
\usepackage{color} 

\begin{document}

\title
{
CCM Calculations For The Ground-State Properties Of 
The One-Dimensional Spin-Half $J_1$--$J_2$ Model: \\
Possible Evidence of Collinear Ordering for $J_2/J_1 > \frac 12$?
}

\author
{
Damian J.J. Farnell
}
\affiliation
{
Division of Mathematics and Statistics, Faculty of Advanced Technology, 
University of Glamorgan, Pontypridd CF37 1DL, Wales, United Kingdom 
}

\date{\today}

\begin{abstract}
In this article we investigate the linear-chain spin-half $J_1$--$J_2$ 
model by using high-order coupled cluster method (CCM) calculations. 
We employ three model states, namely, a nearest-neighbour (n.n.) 
N\'eel model state in which neighbouring spins are antiparallel,
a next-nearest-neighbour (n.n.n.) N\'eel model state in which next-neighbouring 
spins are antiparallel, and finally a type of ``double spiral'' model
state with {\it two} pitch angles. For $J_2/J_1 \le \frac 12$, we find that
the n.n. N\'eel model state produces the lowest energies. 
For $J_2/J_1 > \frac 12$, we find that the stable states for the
quantum system are those for the ``traditional'' spiral state
in which the two pitch angles are identical and the collinear
n.n.n. N\'eel model state. As seen previously, we show that we are 
able to reproduce exactly the dimerised ground (ket) state at the 
Majumdar-Ghosh point ($J_2/J_1=\frac 12$) using the n.n. N\'eel 
model state. We show that the onset of the dimerised phase is 
indicated by a bifurcation of the nearest-neighbour ket- and bra-state
correlation coefficients for the nearest-neighbour N\'eel model state. 
Furthermore, we show that the n.n.n. N\'eel model state can also 
reproduce exactly the dimerized ground (ket) state at the 
Majumdar-Ghosh point ($J_2/J_1=\frac 12$). The ground-state
energies for the n.n.n. N\'eel model state are shown to be 
much lower than those of the ``spiral'' model state for a finite region
above $J_2/J_1=\frac 12$.  Indeed, we show that ground-state 
energies for the collinear n.n. and n.n.n. model states are in 
good agreement with the results of exact diagonalisations  
of finite-length chains across this entire regime for $J_1>0$. 
We produce results for the dimer order parameter for  $J_2/J_1 > 
\frac 12$ by using the n.n.n. N\'eel model state and these
results are in reasonable agreement with previous results 
of DMRG. Finally, results for the spin-spin correlation 
functions for the n.n.n. N\'eel model state are presented.
It is shown that the correlation length increases with increasing
$J_2/J_1$ above $J_2/J_1=\frac 12$ using this model
state and that shift in the peak of the structure function
occurs from $q=\pi$ to $q=\pi/2$ as one increases
$J_2/J_1$. Although these results are not conclusive, 
they are intriguing because they suggest that a ground state exhibiting 
collinear order might occur for some region for $J_2/J_1 \ge \frac 12$. 
\end{abstract}

\maketitle

\section{Introduction}

The formation of dimer- and plaquette-ordered singlet 
ground states (so-called valence-bond crystal (VBC) states) 
is an interesting and important topic in quantum spin systems.
Often, the formation of enhanced dimer or plaquette correlations 
is driven by frustration, which 
can increase quantum fluctuations and which may 
result in such gapped 
rotationally-invariant quantum paramagnetic states.
An example for the appearance 
of such exact VBC product eigenstates are the 
spin-half $J_1$--$J_2$ model on the linear chain [1-12]
at the point $J_2/J_1=\frac 12$ (the 
 so-called Majumdar-Ghosh
point), which is an example of spontaneous symmetry breaking. 
 
\begin{figure}
\epsfxsize=13cm
\centerline{\epsffile{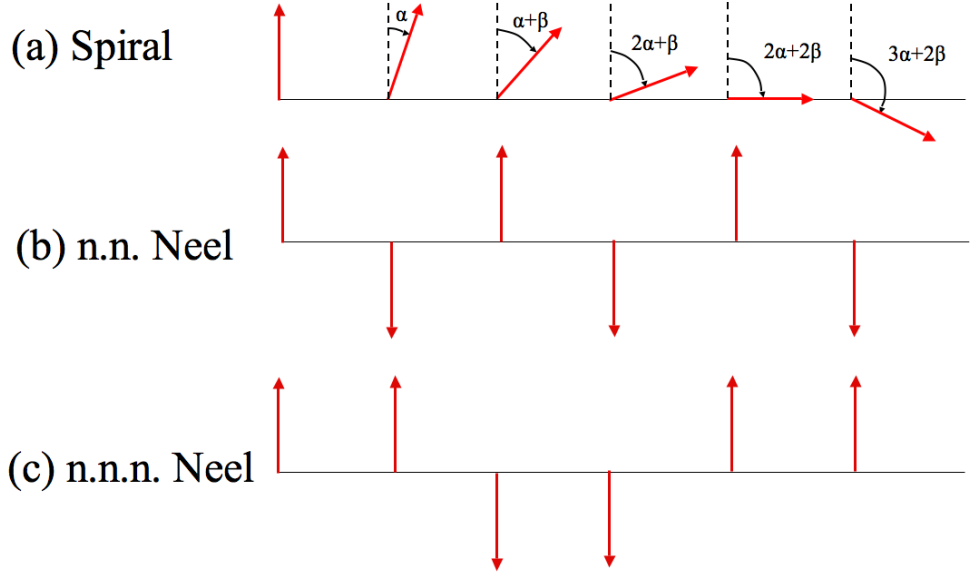}}
\caption{Model states for the one-dimensional spin-half $J_1$--$J_2$ model. 
(a) Spiral model states with two angles; (b) collinear nearest-neighbour (n.n.) 
N\'eel model state; and, (c) collinear next-nearest-neighbour (n.n.n.) N\'eel 
model state}
\label{fig1}
\end{figure}
 
The Hamiltonian for this spin-half model has nearest-neighbour (n.n.) bonds of 
strength $J_1$ and next-nearest-neighbour (n.n.n.) bonds of strength $J_2$ is given by
\begin{equation}
H = \frac {J_1}2 \sum_{ i,\rho_1 } {\bf s}_i \cdot {\bf s}_{i+\rho_1} 
    +    \frac {J_2}2 \sum_{ i,\rho_2 }  {\bf s}_i \cdot {\bf s}_{i+\rho_2} ~~ ,
\label{j1j2H}
\end{equation}
where the index $i$ runs over sites on the lattice, $\rho_1$ runs 
over all  nearest-neighbours to $i$, and $\rho_2$ runs over all  
next-nearest-neighbours to $i$. Henceforth we put $|J_1|=1$  and 
consider $J_2 \ge 0$.

The ground-state properties of this system have been studied 
using methods such as exact diagonalizations  [2,7],
DMRG  [3-5,9], 
CCM [8-11],
and field-theoretical approaches [5]
(see Refs. [5,6] 
for a general review).  At $J_2/J_1=0$ we have the 
unfrustrated Heisenberg antiferromagnet, where the exact solution is provided by the 
Bethe Ansatz. The ground state is gapless and the spin-spin 
correlation function $\langle  {\bf s}_i  \cdot {\bf s}_j\rangle$  decays slowly to zero 
according to a power-law, i.e. no true N\'eel-like long-range order is observed. 
In the region $J_2/J_1>0$ the nearest-neighbour ($J_1$) and next-nearest-neighbour
interactions ($J_2$) compete, thus leading to frustration. At $J_2/J_1 = 0.2411(1)$
the model exhibits a transition to a two-fold degenerate gapped dimerized 
phase with an exponential decay of the correlation function $\langle  {\bf s}_i \cdot 
{\bf s}_j\rangle$ [2-6].
This state breaks the translational 
lattice symmetry. At the Majumdar-Ghosh point $J_2/J_1=\frac 12$ 
there are two degenerate   simple exact dimer-singlet product ground 
states corresponding to the dimerized product state 
for the Hamiltonian of Eq. (\ref{j1j2H}) [1].
The ground state for $J_2/J_1>\frac 12$ is not completely well-understood,
although it is known that gap [5]
persists for some range above
$J_2/J_1=\frac 12$ and that incommensurate spiral correlation occur. 
Furthermore, DMRG studies suggest that a Liftschitz point occurs at 
$J_2/J_1=0.52$ and that the dimer order parameter peaks at $J_2/J_1=0.58$.
In the limit $J_2/J_1 \rightarrow \infty$, the system decouples into two
Bethe chains and so the system is gapless.

\section{Method}

\subsection{The CCM  Formalism}

A description of the underlying methodology for the CCM is given in Refs. [13-21].
We note here that the 
ket and bra ground-state energy eigenvectors, $|\Psi\rangle$ and 
$\langle\tilde{\Psi}|$, of a general many-body system described by a Hamiltonian $H$
\begin{equation} H |\Psi\rangle = E_g |\Psi\rangle\;; \;\;\;  \langle\tilde{\Psi}| H = E_g \langle\tilde{\Psi}| \;, \label{ccm_eq1} 
\end{equation} 
are parametrised within the single-reference CCM as follows:   
\begin{eqnarray} |\Psi\rangle = {\rm e}^S |\Phi\rangle \; &;&  \;\;\; S=\sum_{I \neq 0} {\cal S}_I C_I^{+}  \nonumber \; , \\ \langle\tilde{\Psi}| = \langle\Phi| \tilde{S} {\rm e}^{-S} \; &;& \;\;\; 
\tilde{S} =1 + \sum_{I \neq 0} \tilde{{\cal S}}_I C_I^{-} \; .  \label{ccm_eq2} \end{eqnarray} 
Note that  $|\Phi\rangle$ is the normalised single model or reference state.
The correlation operator $S$ is thus a linked-cluster operator that is decomposed in Eq. (\ref{ccm_eq2}) in terms of a complete set of mutually commuting multi-spin and multi-configurational creations operators $C_I^{+}$ with respect to the model state. The hermitian adjoints, $C_I^{-}=(C_I^{+})^{\dag}$ are the corresponding destruction operators, $\langle \Phi | C_I^{+} = 0 = C_I^{-} | \Phi \rangle$, and we explicitly  define $C_0^{+}=\rm{1}$, the identity operator. The label $I$ is thus a set-index comprising a set of single-particle labels in some suitable single-particle basis defined via $|\Phi\rangle$. We note that the normalisation conditions $\langle \Phi | \Psi \rangle =  \langle \Phi | \Phi \rangle = 1 = \langle \tilde \Psi | \Psi \rangle$ follow from Eq. (\ref{ccm_eq2}). The ground ket- and bra-state equations are given in terms of $\bar{H}=\langle \tilde \Psi | H | \Psi \rangle$ as
\begin{eqnarray} \delta{\bar{H}} / \delta{\tilde{{\cal S}}_I} =0 & \Rightarrow &   \langle\Phi|C_I^{-} {\rm e}^{-S} H {\rm e}^S|\Phi\rangle = 0 ,  ~ \forall ~ I \neq 0 \;\; ; \label{ccm_eq7} \\ \delta{\bar{H}} / \delta{{\cal S}_I} =0 & \Rightarrow & \langle\Phi|\tilde{S} {\rm e}^{-S} [H,C_I^{+}] {\rm e}^S|\Phi\rangle = 0 , ~ \forall ~ I \neq 0 \; . \label{ccm_eq8}
\end{eqnarray}
In order to solve for the one-dimensional $J_1$--$J_2$ model considered here, we make approximations in both $S$ and $\tilde S$.  The three most commonly employed approximation schemes previously utilised have been: (1) the SUB$n$ scheme, in which all correlations involving only $n$ or fewer spins are retained, but no further restriction is made concerning their spatial separation on the lattice; (2) the SUB$n$-$m$  sub-approximation, in which all SUB$n$ correlations spanning a range of no more than $m$ adjacent lattice sites are retained; and (3) the localised LSUB$m$ scheme, in which all multi-spin correlations over all distinct locales on the lattice defined by $m$ or fewer contiguous sites are retained. 
Equation (\ref{ccm_eq7}) shows that the ground-state energy at the stationary point has the simple form 
\begin{equation} 
E_g = E_g ( \{{\cal S}_I\} ) = \langle\Phi| {\rm e}^{-S} H {\rm e}^S|\Phi\rangle\;\; . \label{ccm_eq9}
\end{equation}  

In this article we focus on the application of high-order (CCM) [22-24] 
to the spin-half, one-dimensional $J_1$--$J_2$ model.
The CCM calculations were carried out to high-order using a 
computational code written by us [25].
The first CCM analyses [8,9]
concentrated the N\'eel model state with neighbouring spins
antiparallel [8]
and then for a spiral model state
[8,9] in which there is a pitch angle 
between neighbouring spins. It was assumed that translational 
symmetry was preserved in both cases, and these calculations
led to good results for the ground-state energies of this system.
In Refs. [10,11], 
we used a ``doubled'' unit cell including 
two neighbouring sites for this spin-half system on the linear chain at 
points (0,0,0) and (1,0,0) and a single Bravais vector 
(2,0,0)$^T$ to take into account the symmetry breaking. 
There are thus two distinct types of two-spin nearest-neighbor 
ket-state correlation coefficients, namely, those connecting the 
sites inside the unit cell and those connecting different unit cells. 
These  coefficients are denoted as ${\cal S}_2^a$ and ${\cal S}_2^b$, 
and it is straightforward \cite{ccm_mg3,ccm_mg4} to prove that 
the ground state at $J_2/J_1=\frac 12$ is obtained {\it exactly} by setting 
${\cal S}_2^a=1$ and all other coefficients equal to zero.  
Interestingly, a similar exact ground state at $J_2/J_1=\frac 12$
may also be formed for the n.n.n. N\'eel model state and 
again ${\cal S}_2^a=1$ and all other correlation coefficients 
are zero.

We use three model states, namely, a spiral model state, 
and n.n. and n.n.n. N\'eel model states, shown in Fig. \ref{fig1}. 
For all of these model states, we rotate the spin coordinates 
of the `up' spins so that notationally they become `down' 
spins in these locally defined axes. For example, the relevant 
Hamiltonian in these rotated coordinates for the model state (a)
is then given by
\begin{eqnarray}
H &=& \sum_{ \langle i \rightarrow j \rangle } J_{i,j} \bigl \{
          \cos(\theta_{i,j})     s_i^z s_j^z + 
\frac 14 (\cos(\theta_{i,j})+1) (s_i^- s_j^+ + s_i^+ s_j^- ) 
\nonumber \\
  &&
+\frac 14 (\cos(\theta_{i,j})-1) (s_i^+ s_j^+ + s_i^- s_j^- )+
\frac 12  \sin(\theta_{i,j})    ((s_i^-s_i^+)s_j^z - s_i^z(s_j^-+s_j^+))
\bigr \}
\label{j1j2Hrotated}
\end{eqnarray}
where $\theta_{i,j}$ is the difference between pitch angles at 
sites $i$ and $j$. Furthermore, we see for model state (a) that
$\theta_{2i,2i+1}=\alpha$, $\theta_{2i+1,2i+2}=\beta$, and 
$\theta_{i,i+2}=\alpha+\beta$ where $i=0,1,2,\cdots$. 
(We note also that $J_{i,i+1}=J_1$ and $J_{i,i+2}=J_2$.) 
We remark that both model states (b) may be obtained from model 
state (a) by setting $\alpha=\beta=\pi$ and model state (c) by 
$\alpha=\pi$ and $\beta=0$. Model state (a) allows us to 
investigate the stability of the collinear n.n. and n.n.n. 
model states, as well as the special case $\alpha=\beta$  
previously considered in Refs. [8,9].

\section{Results}

\begin{figure}
\epsfxsize=13cm
\centerline{\epsffile{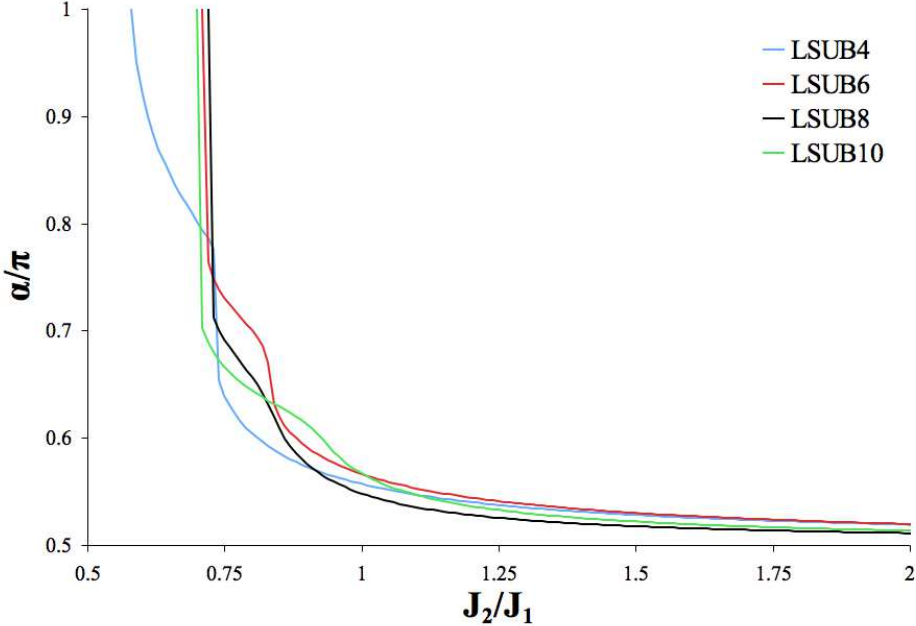}}
\caption{Results for the pitch angle $\alpha$(=$\beta$) versus $J_2/J_1$
for the spiral model state (a).}
\label{fig2}
\end{figure}

For $J_2/J1\le\frac 12$, the n.n. N\'eel state is found to yield the best results.
For $J_2/J1\ge\frac 12$, there were found to be two stable solutions with
respect to the ``pitch'' angles $\alpha$ and $\beta$, namely: the collinear
n.n.n. N\'eel model state ($\alpha=\pi$ and $\beta=0$ or vice versa), which 
breaks the symmetry of the lattice; and, a  spiral state with $\alpha=\beta$, 
which therefore does not break the translational symmetry of the lattice.
We may plot the angle $\alpha$(=$\beta$) for the spiral solution, and this is
shown in Fig. \ref{fig2}. We see that the spiral solution is ``lost'' at around
$J_2/J_1\approx 0.7$, as noted previously in Ref. [9].
There is a direct and finite step-change to $\alpha=\pi$ at this point. Furthermore, 
we see slight oscillations in $\alpha$ for $J_2/J_1 \lesssim 1$.  It is unclear 
why this should occur, although perhaps this behaviour may indicate some 
rapid change in the ground state and/or that the model state is becoming 
increasingly inappropriate.

\begin{figure}
\epsfxsize=13cm
\centerline{\epsffile{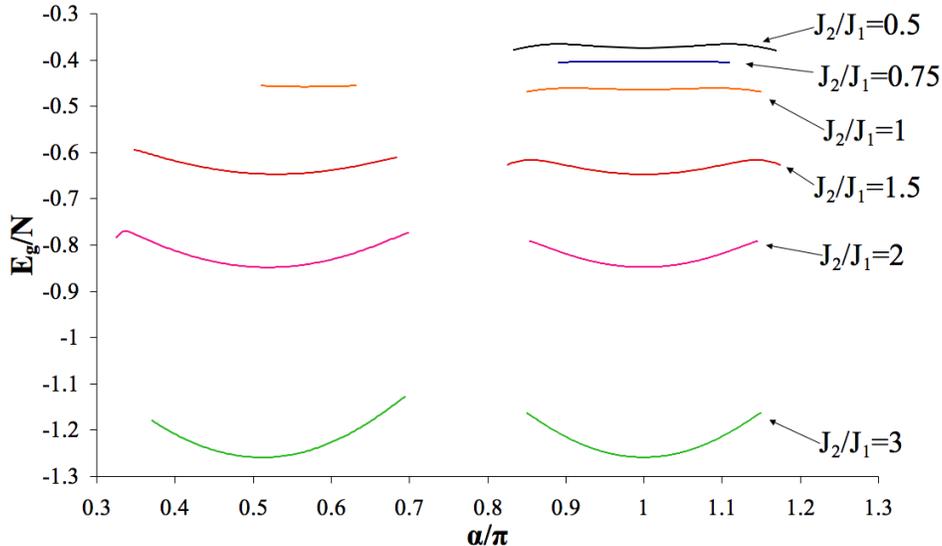}}
\caption{Results for the CCM ground-state energy versus pitch angle
$\alpha$ as a function of $J_2/J_1$ at the LSUB6 level of approximation. 
Results on the right of the figure are for $\beta=0$ (so are for the n.n.n. 
N\'eel model state (c)) and those on the left are for $\beta$ given in Fig. 
\ref{fig2} are so are for the spiral case.}
\label{fig3}
\end{figure}

We may investigate these solutions further by fixing the value of $\beta$ 
and then varying $\alpha$, see Fig. \ref{fig3}. For the n.n.n. N\'eel model
state, we may set $\beta=0$ for all values of $J_2/J_1$ and we see from this 
figure that the minimum energy solution is given by $\alpha=\pi$ in this case
(i.e., identically model state (c)). 
Furthermore, we may set the value of $\beta$ to be given by the value shown 
in Fig. \ref{fig2} (at a given level of LSUB$m$ approximation and for a 
given value of $J_2/J_1$) in order to investigate the ``spiral'' case. 
We note  that the stable minimal solution is given by $\alpha=\beta$, which
preserves the translational symmetry of the lattice. Note that no other
intermediate solutions, e.g., a kind of double spiral in which $\beta=0$
and $\alpha\ne \pi$ were seen to occur. For $J_2/J_1>\frac 12$, there was 
either the translational symmetry-breaking case of model state (c) or the 
spiral state with $\alpha=\beta$, which preserves translational symmetry.

\begin{figure}
\epsfxsize=13cm
\centerline{\epsffile{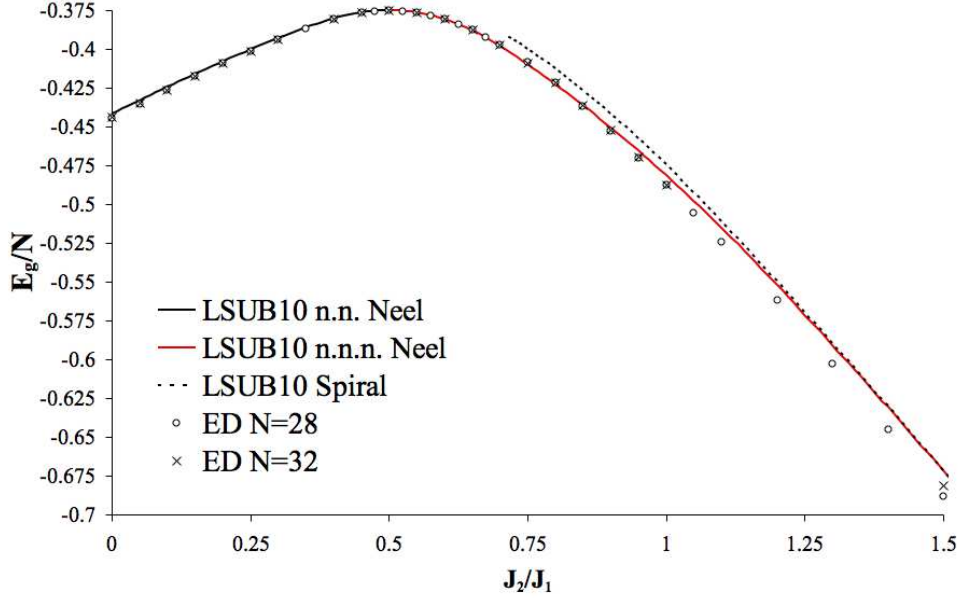}}
\caption{Results for the CCM ground-state energy as a function of $J_2/J_1$
for the collinear n.n. N\'eel model state (b) (for $J_2/J_1\le\frac 12$) and n.n.n. N\'eel 
model state (c) (for $J_2/J_1\ge\frac 12$)  and the spiral model state (a) with 
$\alpha=\beta$.}
\label{fig4}
\end{figure}

We remark that the n.n. N\'eel model state (b) (i.e., $\alpha=\beta=\pi$) 
generally better produces for $J_2/J_1 \le \frac 12$ than the n.n.n. N\'eel 
model state (c), and vice versa in the region $J_2/J_1 \ge \frac 12$. 
Results for the ground-state energies as a function of $J_2/J_1$ are shown 
in Fig. \ref{fig4}. Of the collinear model states (b) and (c), results for the 
n.n. N\'eel model state (b) are shown for $J_2/J_1<\frac 12$ and those of 
the n.n.n. N\'eel model and (c) for $J_2/J_1>\frac 12$. Exactly at the MG point 
$J_2/J_1=\frac 12$, we find that these two states have equal energy.  
Furthermore, this solution is the CCM symmetry-broken solution (more is 
said of this below), which forms an exact ket eigenstate at this point.
For $J_2/J_1\ge \frac 12$, we find that the $\alpha=\pi$ and $\beta=0$ (i.e., model 
state (c)) is lowest until  $J_2/J_1|_{c_2}$ equal to 1.02, 1.69, 1.29, and 1.76 
at the LSUB4 to LSUB10 levels of approximation respectively. (The 
non-monotonic behaviour of this crossing is probably due to a ``double'' odd/even 
effect, where the ``doubled'' comes from the range of the $J_2$ bonds.)
Above this point, the energies for the spiral model state (a) with $\alpha=\beta$
appear to be lower. Symmetry breaking might therefore persist 
for some region for $J_2/J_1\ge \frac 12$. 
The differences between the energies of these two states above this
crossover point are very small however,  e.g., it is of order $10^{-4}$ 
at the LSUB10 approximation. The difference only becomes large for
$J_2/J_1<J_2/J_1|_{c_2}$, where the symmetry-breaking n.n.n. N\'eel collinear 
model state (c) is clearly much lower than the spiral solution for $\alpha=\beta$. 
Furthermore, it is possible that the crossover goes to higher and higher values 
of $J_2$ with increasing level of LSUB$m$ approximation level. However, 
we cannot rule out the possibility that another state of more ``exotic''
order that is not accessible from the model states used here might occur just 
above $J_2/J_1=\frac 12$. If so then this state might exhibit lower 
ground-state energies. 

\begin{figure}
\epsfxsize=13cm
\centerline{\epsffile{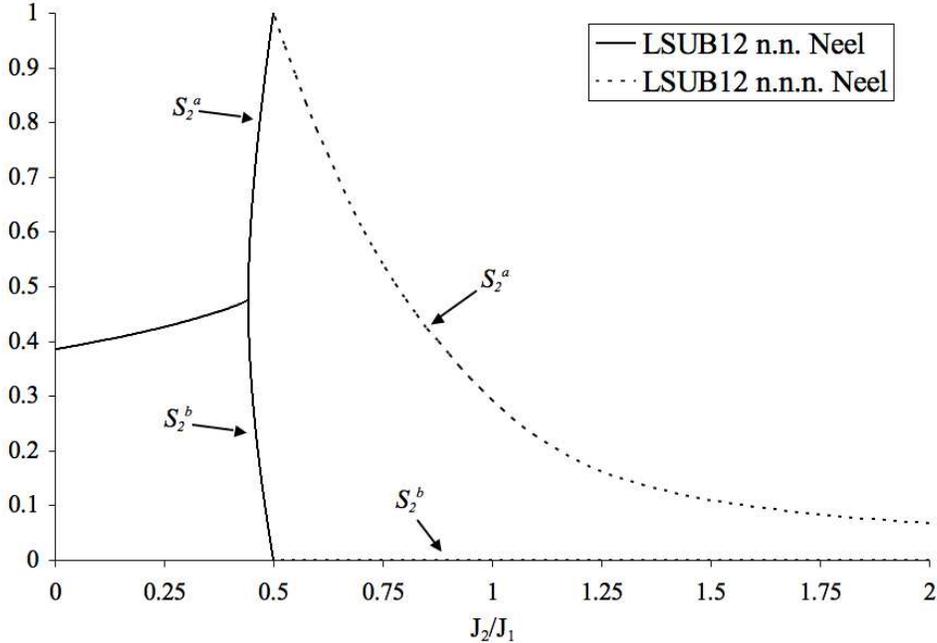}}
\caption{Results for the CCM ket-state correlation coefficients at the 
LSUB12 level of approximation for the one-dimensional spin-half 
$J_1$--$J_2$ model using the collinear nearest-neighbour (n.n.) 
N\'eel model state and the collinear next-nearest-neighbour (n.n.n.) 
N\'eel model state.}
\label{fig5}
\end{figure}

Henceforth, we concentrate on results for the collinear n.n. N\'eel model 
state (b) for $J_2/J_1 \le \frac 12$ and the collinear n.n.n. N\'eel model 
and (c) only for $J_2/J_1 \ge \frac 12$.  Results for the ket-state correlation 
coefficients for the n.n. and n.n.n. N\'eel model states (b) and (c) are 
shown as a function of $J_2/J_1$ in Fig. \ref{fig5}. We remark again
that both model states (b) and (c) provide an exact solution at 
$J_2/J_1=\frac 12$, namely, where ${\cal S}_2^a=0$ and all 
other ket-state correlation coefficients are zero. 
Starting from $J_2/J_1=\frac 12$ we were able to track this exact
solution (as seen previously also in Refs. [10,11])
within a certain LSUB$m$  approximation for the n.n. N\'eel model state to 
lower values of $J_2/J_1 \le \frac 12$. It was also found previously 
[10,11] 
that the solution (i.e. 
${\cal S}_2^a={\cal S}_2^b$) having the full translational symmetry 
[8,9] 
was the only solution below a critical
point $J_2/{J_1}|_{c_1}$($<\frac 12$). Above this point the CCM solution 
for the n.n. correlation coefficient bifurcates, as shown in Fig. \ref{fig5}. 
CCM results for the ground-state for the symmetry-breaking solution 
to the CCM equations for $J_2/{J_1}>J_2/{J_1}|_{c_1}$ compared 
better to exact diagonalisation results in this regime than those 
results for the ``usual n.n. (`N\'eel-type') solution'' that do not
break translation symmetries (see Refs. [10,11] for more details.) 
However, as was also noted before [10,11],
the value for $J_2/{J_1}|_{c_1}$ is much too large (e.g., 
$J_2/{J_1}|_{c_1}=0.436$ at the for LSUB14 level of approximation). 
This value ought to be compared to the value
of $J_2/J_1 = 0.2411(1)$ the model exhibits a transition 
to a two-fold degenerate gapped dimerized phase. 

As we increase the level of approximation the value for $J_2/{J_1}|_{c_1}$ 
does seem to decrease, albeit slowly. In Ref. [11], 
it was proposed that results for the predicted energy gap at a given level 
of LSUB$m$ approximation are overestimated, with good results given 
only after extrapolation. This might artificially stabilise the ``usual CCM 
solution'' tat preserves translational symmetry and that is known to be gapless
Hence  $J_2/{J_1}|_{c_1}$ might be grossly overestimated. However, it is fair 
to say that this proposition is mostly speculation. We believe however 
that the detection of any such ``symmetry-breaking'' solutions via the 
CCM is strong point of the method. 

We note again that the exact solution at $J_2/J_1=\frac 12$ may be
obtained for the n.n.n. N\'eel model state (c) by setting ${\cal S}_2^a=1$ 
and all other coefficients equal to zero. This solution may be tracked for 
$J_2/J_1 > \frac 12$ and the results for the n.n. ket-state correlation 
coefficients are also shown in Fig. \ref{fig5}. We remark again that
this is the minimum energy solution compared to the ``spiral'' solution
up to some possible cross-over value $J_2/J_1|_{c_2}$. Importantly, the 
ket-state correlation coefficient ${\cal S}_2^a$ is non-zero (and 
${\cal S}_2^b$ is zero) for all values of $J_2> \frac 12$ for this model 
state, which suggests that the symmetry-broken ground state might 
persist for some range of $J_2/J_1 (\gtrsim \frac 12)$.

\begin{figure}
\epsfxsize=13cm
\centerline{\epsffile{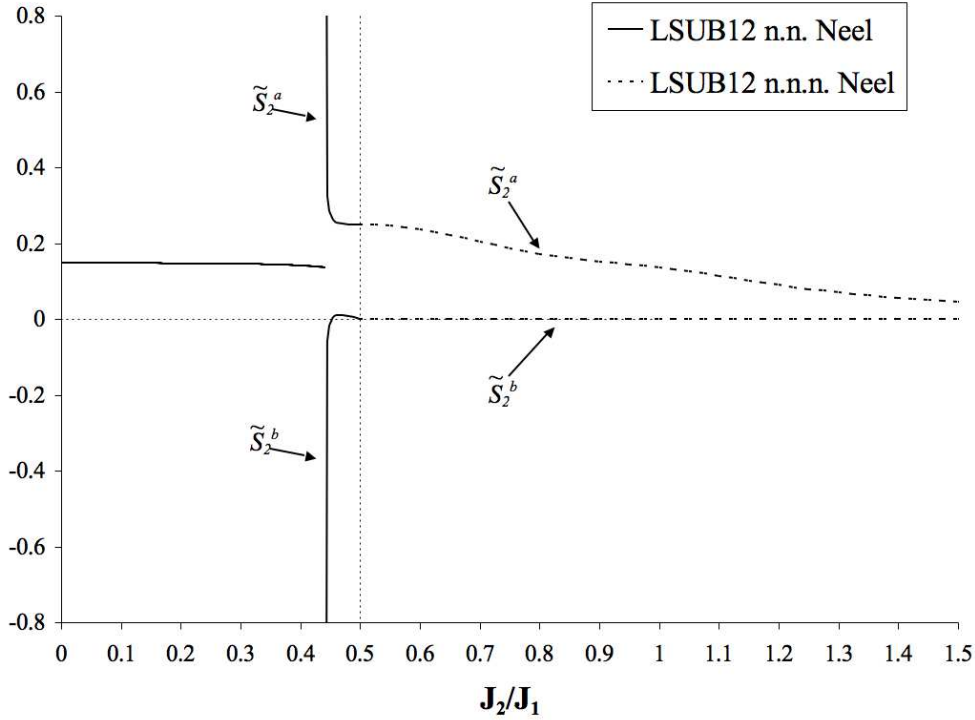}}
\caption{Results for the CCM bra-state correlation coefficients at the 
LSUB12 level of approximation for the one-dimensional spin-half 
$J_1$--$J_2$ model using the collinear nearest-neighbour (n.n.) 
N\'eel model state and the collinear next-nearest-neighbour (n.n.n.) 
N\'eel model state.}
\label{fig6}
\end{figure}

The nearest-neighbor bra-state correlation coefficient at the LSUB$m$ 
level of approximation at $J_2/J_1=\frac 12$ has $\tilde {\cal S}_2^a=1/4$ 
with $m \ge 4$. This is shown in Fig. \ref{fig6} for the LSUB12 level of 
approximation. We find that the nearest-neighbor correlation coefficient diverges 
as $J_2/{J_1} \rightarrow J_2/{J_1}_{c_1}$ and this is also shown in Fig. 
\ref{fig6}. Again, the  usual (`N\'eel-type') solution  ($\tilde {\cal S}_2^a= 
\tilde {\cal S}_2^b$)  is obtained for $J_2/{J_1}<J_2/{J_1}|_{c_1}$. 
Again, a similar solution at $J_2/J_1=\frac 12$ may also be be found for 
n.n.n. N\'eel model state (c), again for which ${\tilde {\cal S}}_2^a=0.25$ 
and all other coefficients equal to zero. In the limit $J_2/{J_1} \rightarrow
\infty$, the bra-state correlation coefficient $\tilde {\cal S}_2^a$ goes
to zero. Once more, qualitatively similar results are observed at other 
levels of LSUB$m$ approximation for the bra-state correlation coefficients 
as a function of $J_2/J_1$.

\begin{figure}
\epsfxsize=13cm
\centerline{\epsffile{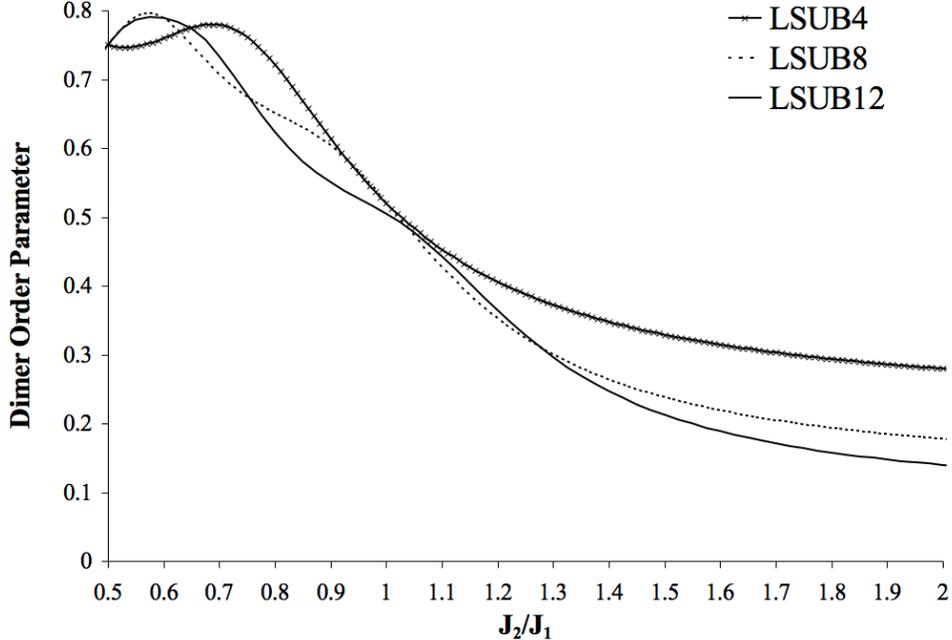}}
\caption{Results for the dimer order parameter for the one-dimensional spin-half 
$J_1$--$J_2$ model using the collinear next-nearest-neighbour (n.n.n.) 
N\'eel model state.}
\label{fig7}
\end{figure}

The dimer order parameter is given by $d=\langle {\bf s}_{2i} \cdot ({\bf s}_{2i+1} 
-{\bf s}_{2i-1} ) \rangle$. We find this quantity for the n.n.n. N\'eel model state
and the results for $J_2>\frac 12$ are shown in Fig. \ref{fig7}. Results
for the n.n. N\'eel model state show an unphysical divergence at the at the 
critical point  $J_2/{J_1}|_{c_1}$, which is due to a divergence in the
NCCM bra-state correlation coefficients as shown above. (This is uninteresting 
and so is not shown here.) We  remark that there is a maximum in $d$ near to  
$J_2/J_1 \approx 0.6$. For example, at the LSUB12 level of approximation, 
there is a maximum value of the dimer order parameter $d$ of
$d=0.790$ at $J_2/J_1=0.58$. This compares well to the results of 
DMRG in Refs. [5,9] 
by visual comparison. In particular,  it was noted in Ref. [9] 
that the maximum in $d$ occurs
at $J_2/J_1=0.58$, in agreement with these new CCM results. 
The results for the n.n.n. N\'eel model state appear to decay to zero
in the limit $J_2/J_1 \rightarrow \infty$, as expected. However, they
do not seem to agree (at given LSUB$m$ level of approximation) 
with the results of DMRG [5] that predict a decay of $d$ given by 
$d=2.283{\rm exp}(-1.622J_2 /J_1)$. 
Furthermore (and somewhat troublingly), the CCM results also show unphysical 
oscillations in $d$ as one varies $J_2/J_1$. The reason for this is unclear, 
although it might hint at strong changes occurring in the in the ground state 
at this point and/or that the n.n.n. N\'eel model state is possibly a poor choice. 
However, it is probably still fair to say that the CCM results for the dimer order 
parameter broadly follow the same pattern as in Ref. [5,9], although 
quantitative agreement for large $J_2/J_1$ is very poor.  Higher
orders of approximation and/or extrapolation might solve this problem.

\begin{figure}
\epsfxsize=13cm
\centerline{\epsffile{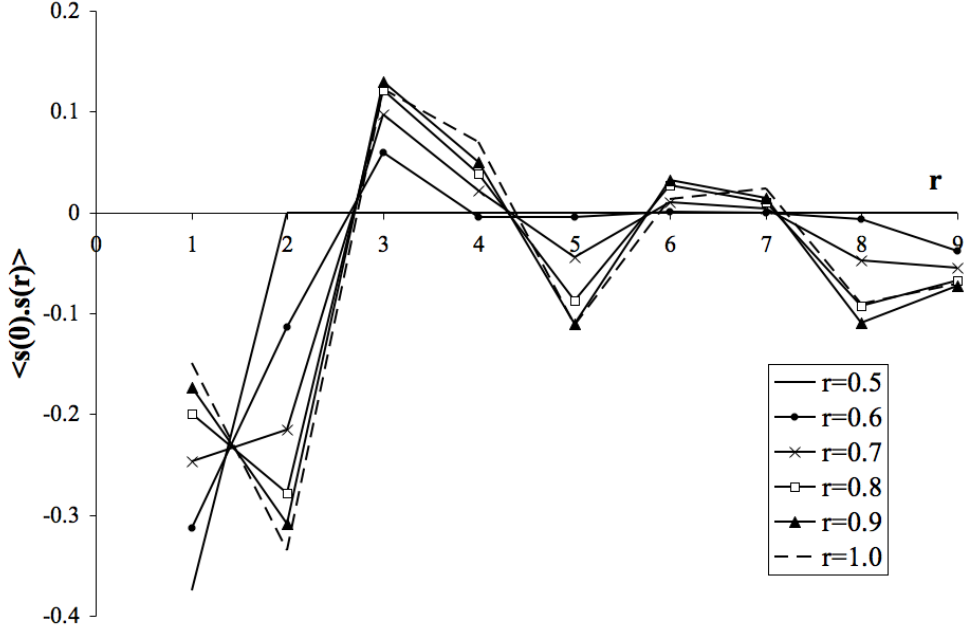}}
\caption{Results for the spin-spin correlation function via the CCM 
for the one-dimensional spin-half $J_1$--$J_2$ model 
using the collinear next-nearest-neighbour (n.n.n.) 
N\'eel model state at the LSUB12 level of approximation.}
\label{fig8}
\end{figure}

\begin{figure}
\epsfxsize=13cm
\centerline{\epsffile{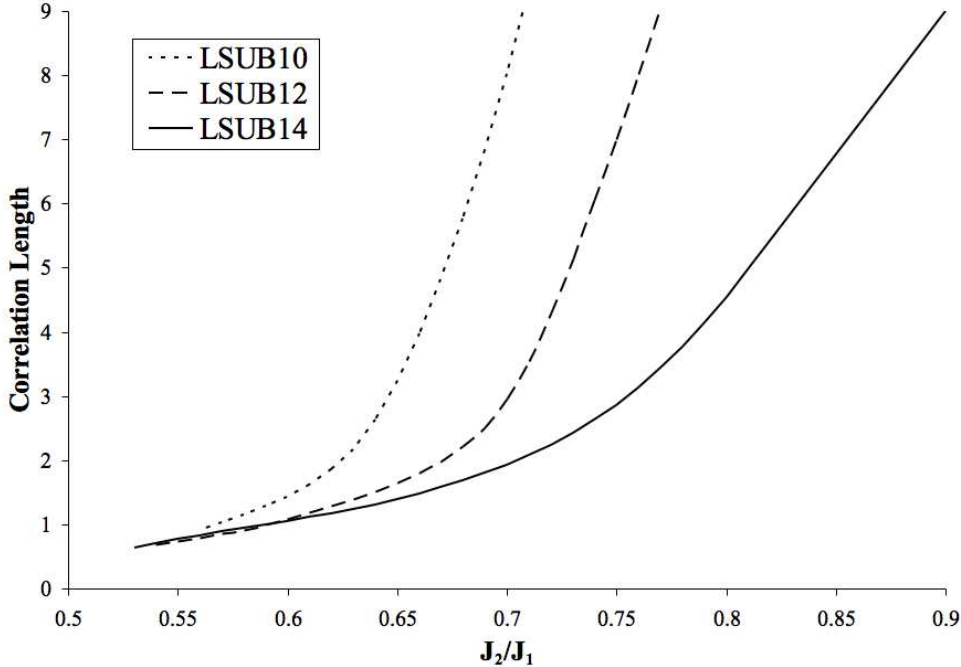}}
\caption{Results for the correlation length via the CCM 
for the one-dimensional spin-half $J_1$--$J_2$ model 
using the collinear next-nearest-neighbour (n.n.n.) 
N\'eel model state at the LSUB10 to LSUB14 levels of approximation.}
\label{fig9}
\end{figure}

Spin-spin correlation functions may be found for the n.n.n. N\'eel model
state (c). At LSUB12 these results are found to be converged to roughly
a separation of $r \approx 10$, shown in Fig. \ref{fig8}, and at LSUB14 they
are  are found to be converged to roughly a separation of $r \approx 12$.
Note however that convergence is also somewhat affected by the value of $J_2/J_1$,
with convergence with increasing level of LSUB$m$ levels of approximation
being quicker at lower values of $J_2/J_1$.
The correlation length $\epsilon$ may be found [5]
by fitting the form $a x^{-0.5} \exp(-x/\epsilon) \cos (\theta x+\lambda)$ 
to the spin-spin correlation functions for the n.n.n. N\'eel model state (c) 
and these results plotted as a function of $J_2/J_1$ in Fig. \ref{fig9}
at the LSUB10 to LSUB14 levels of approximation. By visual inspection
only, these results at the LSUB14 level of approximation do not appear 
to compare too badly to those of DMRG presented in Ref. [5]. 
Indeed, the correlation length does seem to increase with increasing 
$J_2/J_1$. This is an interesting result for a collinear model state of the 
type given by the n.n.n. N\'eel model state (c), although some caution should
be exercised in interpreting these results because we again include 
results for  results only up to $r = 8$ at LSUB10, $r=10$ at LSUB12,
and $r=12$ at LSUB12. We see that the discrepancy between the predicted 
correlation lengths between LSUB10, LSUB12, and LSUB14 becomes larger 
as $J_2/J_1$ increases. However, all levels of approximation
indicate an increase in the correlation length  with increasing 
$J_2/J_1$, as expected.

\begin{figure}
\epsfxsize=13cm
\centerline{\epsffile{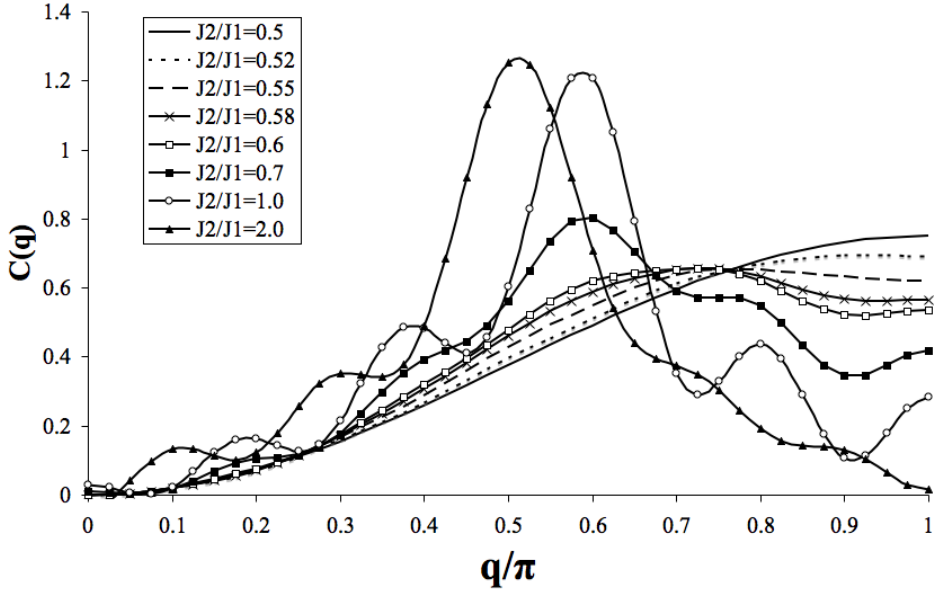}}
\caption{Results for the static structure function $C(q)$ via the CCM 
for the one-dimensional spin-half $J_1$--$J_2$ model 
using the collinear next-nearest-neighbour (n.n.n.) 
N\'eel model state at the LSUB14 level of approximation.}
\label{fig10}
\end{figure}

The static structure function may be defined by
\begin{equation}
C(q)=\frac 12 \sum_{r=-\infty}^{r=+\infty} \langle s({\bf 0}) \cdot s({\bf r}) \rangle
e^{{\rm i}  {\bf r} \cdot {\bf q}}
\end{equation}
However, again our localised LSUB$m$ approximation will function correctly
only for la restricted range of $r$. Results for the static structure function 
at LSUB14 are shown in Fig. \ref{fig9}. We see that the peak of the structure 
function moves from $q=\pi$ at $J_2/J_1=0.5$ to $q=\pi/2$ for larger 
values of $J_2/J_1$. This is in reasonable agreement with previous results of 
Ref. \cite{ccm_mg2} that shows a similar shift in the peak, which 
was interpreted as indicating spiral order. However, the present results
indicate that such a change can occur for a model state with collinear n.n.n.  
ordering also. By visual comparison, the heights of the peaks shown in Fig. \ref{fig10} 
are in reasonable qualitative agreement with those shown in Fig. 4 of Ref. [9]. 
We remark that characteristic and strong oscillations are seen at the 
LSUB14 level of approximation for $J_2/J_1 \approx 1$. This might be 
the model state is a poor choice in this regime, although it is more 
likely that it is because we need to truncate the range $r$ for the LSUB$m$
scheme. Indeed, similar oscillations are seen in Fig. 4 of Ref. [9] and these 
are most probably due to finite-size effects.

\section{Conclusions}

In this article, CCM results were presented for the spin-half 1D $J_1$--$J_2$ 
model. We saw that the n.n. N\'eel model state provided good results for $J_2/J_1
\le \frac 12$. In particular, a translational symmetry-breaking solution was
shown to provide good results for the symmetry-broken phase for this 
model that exhibits a gap (see also Refs. [10,11]). However, as remarked previously 
elsewhere [10,11]
the value for the onset of this phase $J_2/J_1|_{c_1}=0.2411$ was shown 
to be overestimated by this approach.
The stability of the (non-collinear) spiral and (collinear) n.n.n. N\'eel model
states were examined by using a model state that could interpolate between 
the two model states (and incidentally also the n.n. N\'eel model state). It was 
found for $J_2/J_1 \ge \frac 12$ that the only stable solutions were for the case 
of the ``non-translational-symmetry-breaking'' spiral and the (symmetry-breaking) 
n.n.n. N\'eel model state. All other intermediate cases seemed not to be stable. 
Interestingly, the ground-state energies for these two states were extremely close 
for large range of $J_2/J_1$. However, the spiral case appeared to be slightly 
lower in the limit of large values of $J_2/J_1$. Even more startlingly, however, 
the energies for the collinear n.n.n. N\'eel model state became lower than for the spiral 
case in the regime $\frac 12 \lesssim J_2/J_1 \lesssim 2$. This hints at a possible 
``order-from-disorder'' phenomena in this model in this regime. However, these 
results should be treated with caution because the ``true'' ground state in this 
regime might still be quite different to any of the states studied here.

Results for the dimer order parameter $d$ using the n.n.n. N\'eel 
model state showed a maximum 
at $d$ at about $J_2/J_1 \approx 0.58$ (of $d \approx 0.79$ at the LSUB12 
level of approximation), which was in good agreement with DMRG results of Ref. [9].
The decay of $d$ with increasing $J_2/J_1$ did not agree with DMRG
results and also the CCM results also demonstrated ``oscillations'' 
in $d$ as one varied $J_2/J_1$. The n.n.n. nature of the $J_2$ bond means 
that the length scale is effectively doubled and so we probably need to go to 
much higher levels of LSUB$m$ approximation via massive parallel
processing. In order to achieve good quantitative agreement for large $J_2/J_1$, 
we might also need to extrapolate our LSUB$m$ results in the limit $m \rightarrow \infty$ 
However, it is probably still fair to say that the CCM results for the dimer 
order parameter $d$ showed here followed a broadly similar pattern to results 
of DMRG of Refs. [5,9],

The presence of collinear n.n.n. N\'eel ordering some region for $J_2/J_1>
\frac 12$ seems to contradict earlier results of other approximate method that 
indicate incommensurate spiral correlations commonly (see, e.g., Ref. [9]), i.e.,
possibly suggesting  some form of ``spiral'' ground state. Spiral ordering is 
suggested because the classical spiral ground state must also show 
a similar change position of the peak in the (classical) structure function. However, 
we note that our results the static structure function for the n.n.n. N\'eel model 
(e.g., at the LSUB14 level of approximation in Fig. \ref{fig10}) state also showed a 
distinct peak that moved from $q=\pi$ at $J_2/J_1=\frac 12$ to $q=\pi/2$ as 
$J_2/J_1=\rightarrow \infty$. This was also seen in DMRG studies [9] of the quantum 
system. Furthermore, CCM results here predict that the correlation length becomes 
larger as one increases $J_2/J_1$ in the regime $J_2/J_1>\frac 12$,
again as was also seen in DMRG studies [5]. Note however that we used a fairly 
restricted range in $r$ for the spin-spin correlation functions in order to 
form our results for the structure function etc. (only up to a separation of $r \approx 
10,12$ at LSUB14 for the structure function). This fact does not bolster our confidence
in our results, although this approach ought to be adequate in the regime 
just above $J_2/J_1=\frac 12$, where the correlation length is known to 
be very short-ranged. 

In conclusion, the results presented in this article present an intriguing (though 
ultimately inconclusive) possibility that collinear n.n.n. N\'eel ordering might occur 
for $J_2/J_1>\frac 12$. The ground-state energies for the n.n.n. N\'eel model
state are lowest and results for the dimer order parameter, correlation length and 
structure function using this model state are (arguably) predicted correctly. 
If this result is in fact true then it has a number of implications
for the interpretation of similar evidence (e.g., from structure functions) in other
systems where the existence incommensurate ``spiral'' ordering is inferred. 
We suggest that other approximate methods (e.g., spin-wave theory and 
cumulant series expansions) might be applied to the 1D $J_1$--$J_2$ 
model using a n.n.n. N\'eel state as a ``starting point''. We recommend also 
that even higher-order CCM be carried out using massive parallel processing in 
order to confirm the results presented here. 


\end{document}